\documentclass[sigconf]{acmart}
\AtBeginDocument{%
  }

\setcopyright{acmlicensed}
\copyrightyear{2026}
\acmYear{2026}

\acmConference[IndiaHCI '26]{17th Indian Conference on Human-Computer Interaction}
{October 29--31, 2026}{Hyderabad, Telangana, India}
\acmISBN{978-1-4503-XXXX-X/2018/06}

\begin{document}

\title{Beyond the 1D Reality-Virtuality Continuum: A 2D Framework for Seven Extended Reality Paradigms}

\author{Rajashekhar Vachiravelu Saminathan}
\email{raja23@iitk.ac.in}
\orcid{0000-0003-4826-805X}
\affiliation{%
  \institution{HIVE Lab, Department of Design, Indian Institute of Technology - Kanpur}
  \city{Kanpur}
  \state{Uttar Pradesh}
  \country{India}
}

%
%
%
%
%
%

\renewcommand{\shortauthors}{Rajashekhar V S}


\begin{abstract}
The Reality-Virtuality (RV) continuum helps in describing the relationship between physical and computer-generated environments. However, the increasing number of Extended Reality (XR) paradigms introduces new technologies. Their characteristics cannot be completely represented by a single real-virtual dimension. It can be noted that XR systems not only vary in their proportion of real and virtual content but also adds, suppresses, or transforms information available to the user. This work proposes a framework for characterizing seven XR realities, namely Virtual Reality (VR), Augmented Reality (AR), Mixed Reality (MR), Augmented Virtuality (AV), Phenomenal Reality (PR), Annexing Reality (AnR), and Diminished Reality (DR). It uses a two-dimensional extension of the conventional 1D RV continuum. The first dimension helps in identification of the position of an XR technology between real and virtual environments. The second dimension captures information addition and subtraction. The proposed approach gives a qualitative basis for comparing existing and emerging XR paradigms within a unified XR framework.
\end{abstract}



\begin{teaserfigure}
 \includegraphics[width=\textwidth]{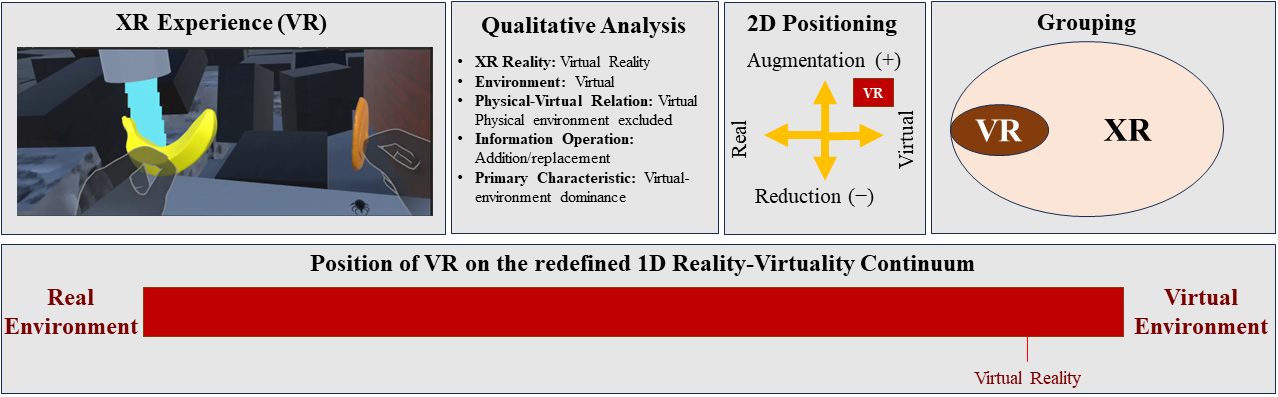}
  \caption{Overview of the work. The XR experience in this case, a VR experience is shown to the participant. A qualitative analysis is then done. Then the 2D positioning of the VR is done which is followed by grouping. This is followed by converting the 2D graph to a 1D Reality-Virtuality continuum.}
  \label{fig:teaser}
\end{teaserfigure}

\received{15 August 2026}
\received[revised]{01 October 2026}
\received[accepted]{10 October 2026}
\maketitle

\section{Introduction}
The field of Extended Reality (XR) covers a diverse set of technologies which is used to modify, mediate, or extend a user's perception of the physical and virtual worlds. These XR technologies traditionally include Virtual Reality (VR), Augmented Reality (AR), Mixed Reality (MR), and Augmented Virtuality (AV) which are commonly described using the Reality-Virtuality (RV) continuum. This continuum gives a conceptual relationship between real and virtual environments \cite{milgram1994taxonomy,milgram1995augmented}. As XR technologies have evolved, the need to fit naturally within a single real-virtual dimension have emerged. These include Phenomenal Reality (PR) \cite{mann2023fundamentals}, Annexing Reality (AnR) \cite{hettiarachchi2016annexing}, and Diminished Reality (DR) \cite{cheng2022towards}. They modify the user's perception and extend beyond increasing or decreasing the amount of virtual content. 

The original RV continuum helps in describing the transition from a physical environment to a virtual environment. However, positioning an XR technology based on its degree of realness or virtuality does not fully describe that technology. It transforms the information available to the user. For example, AR introduces virtual information into the physical environment. On the other hand DR can suppress or remove information that is already physically present. Both of them show different transformations of environmental information, even though they occupy similar regions of the conventional RV continuum. In the recent work there is a question of whether reality and virtuality should be treated as a single continuous dimension. It has called for broader and multidimensional characterizations of XR systems \cite{skarbez2021revisiting,egger2024diminished,han2026reality}.

This work addresses this limitation by proposing a theoretical framework for characterizing and comparing the seven XR realities: VR, AR, MR, AV, PR, AnR, and DR. Our framework extends the conventional 1D RV representation into a two-dimensional space. The horizontal axis represents the position of an XR technology between the real and virtual environments. The vertical dimension represents the direction and magnitude of environmental information manipulation. It ranges from information addition to information subtraction. This helps in finding the XR paradigms that may occupy similar positions on the conventional RV continuum and perform different transformations of the perceived information.

The framework focuses on XR implementations developed for earthquake scenarios. This shows the seven realities across applications including trapped-in-debris experience, evacuation guidance, robotic search, probe usage training, visual clutter reduction, robot-human communication, and robot path finding. The main contributions of this work are threefold:
\begin{enumerate}
    \item A unified characterization of seven XR realities based on a common set of characteristics;
    \item A two-dimensional Reality-Virtuality framework that gives both real-virtual positioning and information about data addition/subtraction; and
    \item A qualitative analysis for positioning XR technologies in the proposed 2D framework and then remapping them in a 1D RV continuum.
\end{enumerate}

The rest of the paper is organized as follows: The Section \ref{sec:backmot} discusses the background and motivation for the work. The Section \ref{sec:relawork} presents the related work. The Section \ref{sec_experiment} shows the experiments done using the 7 XR techniques for earthquake scenarios. The reframing of the 1D RV continuum is presented in Section \ref{sec:reframe}. The preliminary analysis is done in Section \ref{sec:preanalysis}. Then the discussion and limitations  are discussed in Section \ref{sec:dislim}. Finally, the concluding remarks are given in Section \ref{sec:conc}.

\section{Background and Motivation}
\label{sec:backmot}
\subsection{The Reality-Virtuality Continuum}

The relationship between physical and computer-generated environments can be represented using the Reality-Virtuality continuum. Milgram and Kishino positioned ARy and AV between the real and virtual environments. They proposed a taxonomy for describing mixed-reality visual displays \cite{milgram1994taxonomy,milgram1995augmented}. The conventional continuum provides an intuitive one-dimensional representation. The systems closer to the real environment has physical content, whereas systems closer to the virtual environment has computer-generated content. This representation is important since it forms a foundation for describing XR technologies. However, the development of new XR paradigms have limitations in treating realness and virtuality as a single dimension. Skarbez et al. revisited the original continuum and reported that mixed reality is broader than the original formulation \cite{skarbez2021revisiting}. A more recent work has made transition from simple display-oriented classifications toward multidimensional models which consider the system and experiential features \cite{han2026reality}.

\subsection{Beyond the Real-Virtual Dimension}

An XR technique can reproduce real environments or produce virtual scenes. They can add or subtract data or scenes from the display that is shown to the user.  This distinction is important for the seven realities considered in this work. The VR is mostly computer-generated, whereas AR introduces virtual information into a physical environment. The MR integrates real and virtual elements. The AV uses physical-world information and embeds them into a virtual environment. The DR represents suppresses or removes the information from the real world. The AnR extends physical objects or spaces through virtual content. The PR focuses on the realism and experience of content rather dealing with its physical or virtual origin. Here any two technologies can have positions along the real-virtual axis but differ in how they manipulate the environmental information. The conventional continuum does not deal with distinguishing addition, subtraction, and transformation of information.

\subsection{Motivation for a Two-Dimensional Representation}

The motivation for the proposed framework is due to three observations that were made while trying to compare the seven realities. First, the realities differ in physical and virtual content. Second, the XR technologies can add information to the user environment, remove information from it, or transform them. Third, the degree of integration between the physical and virtual elements vary across paradigms. These observations highlight that the seven realities cannot be dealt with using a single dimension. Therefore, this work introduces information manipulation as a second dimension. The horizontal axis retains the conventional real-virtual relationship. The vertical axis represents information augmentation and reduction. The positive (upward) direction indicates increasing information addition or augmentation. The movement toward the negative (downward) direction shows increasing information about subtraction or reduction. This representation makes paradigms such as AR, AnR and DR to be distinguished based on the different ways in which they modify the information presented to the user.

\section{Related Work}
\label{sec:relawork}
The relationship between the physical and virtual worlds has been extensively studied through reality-virtuality (RV) continuum models which was introduces by Milgram and Kishino. They introduced the seminal mixed-reality framework by positioning augmented reality (AR) and augmented virtuality (AV) between the real and virtual environments \cite{milgram1995augmented}. An earlier taxonomy by Milgram et al. classifies mixed-reality visual displays according to the relationship between real and virtual objects \cite{milgram1994taxonomy}. These works established the RV continuum to describe relationships between physical and computer-generated environments. The nature of the RV continuum is questioned whether it can represent the growing diversity of reality-based systems. Skarbez et al. \cite{skarbez2021revisiting} reported that the RV continuum should not be treated as a single continuous dimension. Similarly, a recent work highlights the need for multidimensional frameworks \cite{han2026reality}. These developments show that the position of a technology requires more than the current information of real and virtual content.

The continuum also accounts for transformations of the physical environment beyond the addition of virtual information. Egger et al. \cite{egger2024diminished} showed the removal or diminution of real-world objects. They proposed a directional, layer-based taxonomy that extends beyond conventional mediated and multimediated reality. This is particularly relevant to DR, where there is suppression or removal of information from the physical environment. Mann et al. \cite{mann2018all} proposed a broad taxonomy consisting of virtual, augmented, mixed, mediated, and multimediated realities. The need for clearer boundaries between reality categories has also been identified empirically. McGowin et al. \cite{mcgowin2025reality} did this through a mixed-methods stakeholder study. They found that inconsistent interpretations of VR, AR, and MR create difficulty in communication of the required information. Milman et al. \cite{milman2018defining} contributed to the conceptual clarification of reality categories. They did this by providing distinctions that can be used to identify individual reality paradigms. The research question is: "\textbf{RQ:} \textit{How can existing and emerging XR paradigms be qualitatively characterized beyond their position on the conventional 1D Reality-Virtuality continuum?}

\section{Experimental Exploration of XR Realities in Earthquake Scenarios}
\label{sec_experiment}
\begin{figure*}
\begin{center}
\includegraphics[width=\textwidth]{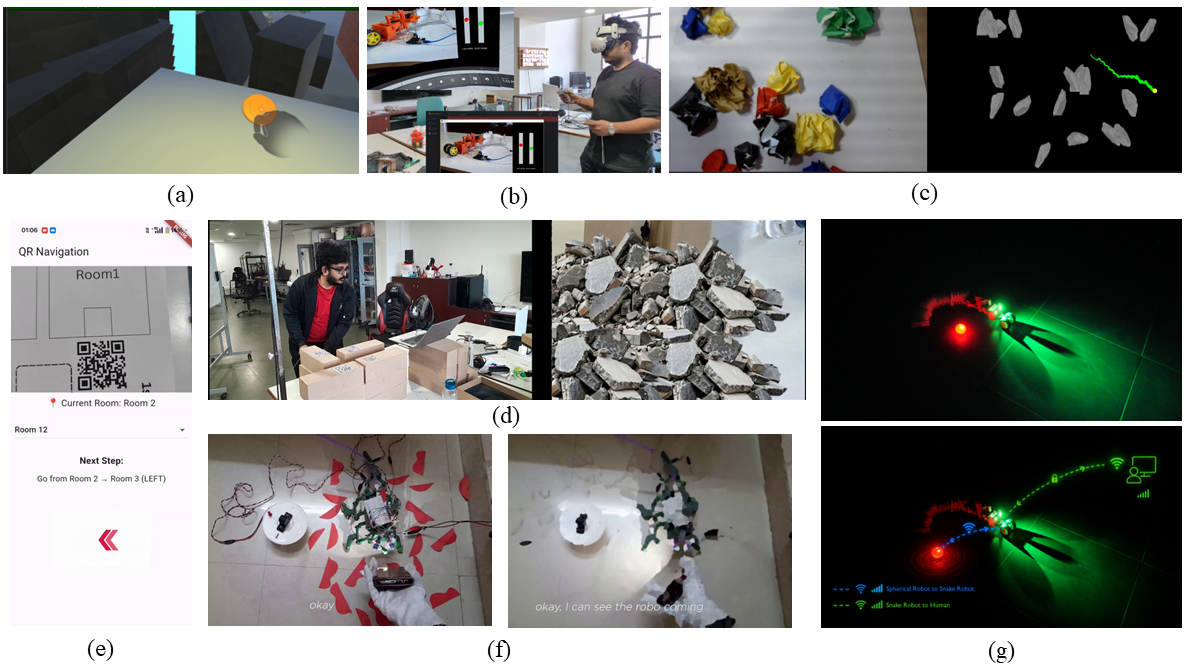} 
\caption{(a) CAVE-E: A scene from the VR experiment (b) MR based soft snake robot control (c) Colour crushed paper used for annexing debris (AnR) (d) The AV experiment with the probe (e) The QR code based evacuation in AR (f) The diminished scenes of the clutter in DR (g) The PR setup that highlights the robot communication.}
\label{fig_experiment}
\end{center}
\end{figure*}

\subsection{Virtual Reality: Trapped in Debris Experience}
The awareness creation about the trapped in debris experience through VR was done using CAVE-E (Controlled Artificial Virtual Earthquake - Environment). The CAVE-E has a virtual environment that is created in the \textit{Unity$^{TM}$} environment. It gives a trapped in debris experience which could be encountered in the post earthquake situations when buildings collapse. More details about this are available in \cite{rajashekhar2025exploring}. A scene in the VR where the user grasps a pancake can be seen in Figure \ref{fig_experiment} (a).

\subsection{Augmented Reality: Evacuation based on QR Codes}
On the occurrence of earthquake, the people in the high raised buildings need to evacuate. This can be done using the AR technique, where the QR codes that are placed on the walls at decision taking pathways help in guiding the people to evacuate. This is done using \textit{A-STAR} algorithm where the user scans the QR code using mobile application for obtaining the direction route. A screen shot is shown in Figure \ref{fig_experiment} (e). 

\subsection{Mixed Reality: Controlling a Dual Soft Robot for Search Operations}
The need for robots for aiding in search and rescue operations is important as discussed in \cite{rajashekhar2026preparedness}. To make the human interaction with the robots immersively, the field of MR plays an important role. A scene where the user operates a dual soft manipulator and snake robot \cite{rajesh2024serpentine} through a \textit{Meta Quest 3} heads-up display is seen in Figure \ref{fig_experiment} (b).

\subsection{Augmented Virtuality: Training for Using the Proposed Probe}
The training of human operators to use passive probes can be done using the AV technique. Here, there are QR codes placed on the wooden boxes which acts as a guide or structured obstacle. The operators view the screen where the camera view of the setup helps in converting the QR codes into immersive (virtual) debris that appear on the screen. This is AV since there are physical probes that are inserted in to the virtual environment and their feedback is felt by the operator. The operator feels that they are inserting the probe in to a realistic debris. Since the debris are dusty and are not easy to carry in to laboratory environments the AV technique can be used. It can be seen in Figure \ref{fig_experiment} (d).

\subsection{Diminished Reality: Reducing Visual Clutter on the Disaster Dashboards}
In the era of perceiving more, it is equally important to perceive less, since it decreases the cognitive load. This act of perceiving less can be done using DR where objects of disinterest or uncomfortable nature can be diminished from the view of the user. There is a robotic lizard \cite{rajashekhar2024optimal} present in a cluttered environment. The scene is diminished where the clutter is removed and inpainted. It is shown in Figure \ref{fig_experiment} (f). 

\subsection{Phenomenal Reality: Communication of Robot Data Transfer Information}
The PR plays a critical role in informing the audience about how they would feel when they are immersed in a situation that is not present in front of them. This can be seen in Figure \ref{fig_experiment} (g) where there are spherical and soft snake robot. The phenomenon of communication among each other and their interaction with human operator can be seen in them.  

\subsection{Annexing Reality: Help in Path Finding of Robots in Debris}
The AnR plays an important role in making the physical object feel real by annexing with other objects that are in the virtual world. In our case, there are crushed colour papers (captured by camera) which depict the virtual debris on the screen. Based on the position of debris, there comes a virtual snake robot that moves from the top left to bottom right of the screen. The snake robot uses \textit{A-STAR} algorithm for finding path inside the virtual debris which appear on their way. This can be seen in Figure \ref{fig_experiment} (c).

\section{Re-Framing the One-Dimensional Reality-Virtuality Continuum}
\label{sec:reframe}
\subsection{Proposed Two-Dimensional Reality-Virtuality Framework}
\begin{figure*}
\begin{center}
\includegraphics[width=\textwidth]{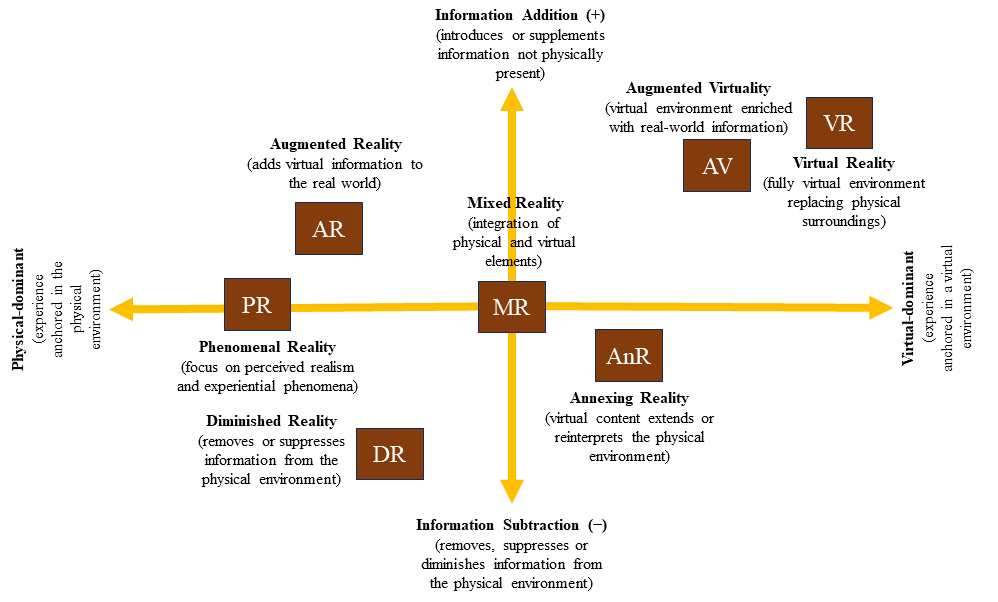}
\includegraphics[width=\textwidth]{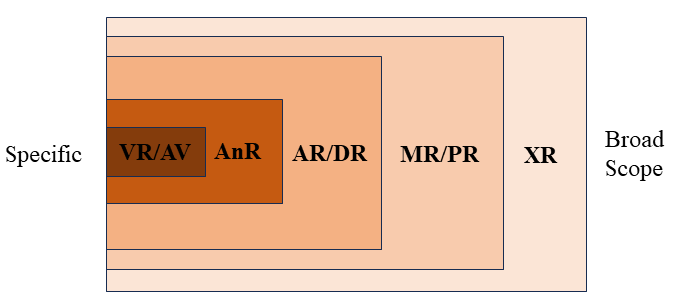} 
\caption{(a) The 2D RV positioning of the seven XR techniques based on the information in Table \ref{tab:qualitative_xr} (b) The grouping of the XR techniques.}
\label{fig_second}
\end{center}
\end{figure*}

The 2D reality-virtuality framework is made  of knowing the position of the XR technology on the real-virtual axis (horizontal axis). On the vertical axis it adds or augments information in the up direction and subtracts or reduces the information in the down direction. We can see that the uncommon XR techniques like PR, DR and AnR have been allotted a position on the 2D graph. This can be seen in Figure \ref{fig_second} (a). 

\subsection{Obtaining an Updated One-Dimensional Reality-Virtuality Continuum}
\begin{figure*}
\begin{center}
\includegraphics[width=\textwidth]{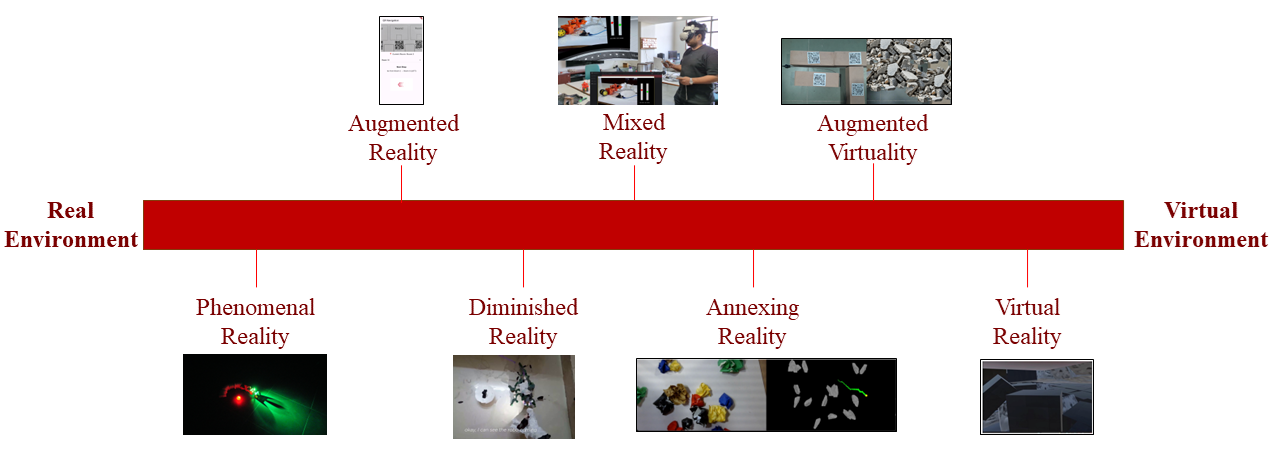} 
\caption{The re-framing of the 1D Reality-Virtuality continuum from the 2D framework.}
\label{fig_positioning_1d}
\end{center}
\end{figure*}

The seven realities that were plotted in Figure \ref{fig_second} are now converted to a re-framed 1D RV Continuum if needed. This is done by moving the XR elements in the 2D space towards the horizontal axis. This gives rise to the updated or re-framed 1D RV Continuum which can be seen in Figure \ref{fig_positioning_1d}. It also has the proposed examples (in Section \ref{sec_experiment}) for the XR technologies which can be seen along with their labelling.

\subsection{Grouping the Seven Realities}
Apart form creating a 2D R-V positioning method, we group the XR technologies as shown in Figure \ref{fig_second} (b). It is  due to the major virtual information involvement, the VR and AV form the inner set. With the presence of significant addition and subtraction of data, AnR is the next set. The property of addition and subtraction of information is common in AR and DR respectively. The MR and PR form the next envelope since there is involvement of real world scenes. This can be seen in Figure \ref{fig_second} (b). If more realities are found in future, they can be added to this group by finding their common nature with the existing XR technologies.

\section{Preliminary Analysis}
\label{sec:preanalysis}
The proposed 2D RV is analysed qualitatively in this section. This was preferred over the quantitative analysis since it is difficult for the author to find participants who have worked with all the seven realities considered here. Hence, a qualitative analysis is done based on the author experience with seven XR realities and presented briefly in the the upcoming section.

\subsection{Qualitative Cross-Paradigm Analysis of the Seven XR Realities}

The seven XR techniques described in Section \ref{sec_experiment} were qualitatively compared based on the author experience. It is done to identify the nature that differentiate the proposed reality paradigms. Instead of treating the realities only as positions on a real-virtual axis, the analysis considers their relationship with the physical environment, virtual content, user interaction, perception, and environmental information. These characteristics are found to provide a qualitative basis for distinguishing the seven realities.

\subsection{Cross-Paradigm Comparison}

The qualitative comparison by the author finds that the seven realities differ in two aspects. It is not only in their relative amounts of physical and virtual content, but also in how environmental information is manipulated and experienced. The VR and AV are mostly virtual environment paradigms. They differ in their treatment of physical-world information. The VR replaces direct access to the physical environment with a computer-generated environment. On the other hand, the AV uses information that originates from the physical world and feeds it to a virtual environment.

The AR and DR have complementary features. They retain the physical environment as the primary reference, but AR adds information whereas DR suppresses the existing information. The MR occupies an intermediate position through the integration of physical and virtual entities. It does interaction across both environments. The AnR introduces virtual information into physical space, but extends or reinterprets the physical objects and spaces using a virtual layer.

The PR differs from the remaining paradigms due to its nature for providing deep insights rather than showing the existing environments. The focus is placed on how a phenomenon is perceived and communicated by the user. On the whole, these observations show that the following conventional 1D real-virtual dimension alone is not sufficient to characterize all seven XR techniques. The analysis identifies two complementary properties: the relationship between physical and virtual environments and the way in which the information is manipulated.

\subsection{Qualitative Method of Positioning the 7 XR Realities}

The resulting qualitative findings are summarized in Table \ref{tab:qualitative_xr}. The table does not assign numerical values to the realities. It identifies the main characteristics observed in each implementation and provides the conceptual basis for the proposed two-dimensional framework. It is shown in Figure \ref{fig_second} (a). The qualitative characterization gives the conceptual basis for extending the conventional 1D RV representation into a 2D framework. The resulting framework allows the less common XR technologies to find a place in the 2D space representation. To claim that every XR implementation belongs exclusively to a single category quantitative analysis needs to be done with a questionnaire and user study with participants who have experienced and built all the 7 XR techniques. 

\begin{table}
\centering
\caption{Qualitative characterization of the seven XR realities.}
\label{tab:qualitative_xr}
\scriptsize
\begin{tabular}{p{0.05\columnwidth} p{0.13\columnwidth} p{0.25\columnwidth} p{0.19\columnwidth} p{0.21\columnwidth}}
\hline
\textbf{Reality} &
\textbf{Environment} &
\textbf{Physical-Virtual Relation} &
\textbf{Information Operation} &
\textbf{Primary Characteristic} \\
\hline

VR &
Virtual &
Physical environment excluded &
Addition / replacement &
Virtual-environment dominance \\

AR &
Physical &
Virtual information supplements reality &
Addition &
Physical environment with virtual annotation \\

MR &
Combined &
Real and virtual elements integrated &
Addition / transformation &
Interaction between real and virtual entities \\

AV &
Virtual &
Physical information incorporated into virtual space &
Addition / incorporation &
Physical-world enrichment of a virtual environment \\

PR &
Perceptual &
Experience-oriented &
Transformation &
Perceived realism and phenomenal experience \\

AnR &
Physical-virtual &
Virtual content extends physical space &
Addition / transformation &
Extension and reinterpretation of physical space \\

DR &
Physical &
Physical information suppressed &
Subtraction &
Reduction of visible environmental information \\

\hline
\end{tabular}
\end{table}

\section{Discussion and Limitations}
\label{sec:dislim}
The proposed framework extends the conventional 1D RV continuum by introducing information manipulation as an additional dimension. The experimental exploration of the seven realities show that XR paradigms differ not only by physical and virtual content but also in how they add or remove information. The 2D representation gives a more expressive description than the conventional 1D RV axis alone. In particular, realities such as AR and DR can be distinguished based on whether information is introduced or removed. Similarly, AnR can be represented according to its combination of real-virtual positioning. The proposed representation allows emerging XR paradigms to be positioned in a way, that is, only by their location on the traditional continuum.

The proposed framework has several limitations. The current work does not establish the generalizability of the 2D RV framework across a large and diverse population. The information addition and subtraction axis (vertical axis) capture the direction and magnitude of information manipulation. It does not fully describe the semantic or cognitive consequences of that manipulation. The seven realities considered in this study represent the current scope of the framework. The present framework is demonstrated using XR implementations developed for earthquake scenarios. Hence, the results may not capture all properties that are relevant in other application domains. Despite these limitations, the framework gives a starting point by qualitatively comparing the 7 XR realities. A structured empirical validation, for example, a card-sorting study or a Delphi study with experts will be done in future.  

\section{Conclusions}
\label{sec:conc}
This work presented a framework for characterizing seven Extended Reality paradigms: Virtual Reality, Augmented Reality, Mixed Reality, Augmented Virtuality, Phenomenal Reality, Annexing Reality, and Diminished Reality. The study was done since we identified a limitation of the conventional 1D Reality-Virtuality continuum. To address this limitation, a 2D RV framework was proposed. The horizontal dimension represents the position of an XR technology between real and virtual environments. The vertical dimension represents information addition and subtraction. The resulting 2D representation also provides a systematic route for mapping the realities back onto a re-framed 1D RV continuum. Future work should validate the framework quantitatively on a larger and more diverse participant groups.  Such studies can further establish the reliability and generalizability of the proposed 2D Reality-Virtuality approach.

\bibliographystyle{ACM-Reference-Format}
\bibliography{sample-base}


\end{document}